\documentclass{llncs}

\newcommand{\citep}[1]{\cite{#1}}
\newcommand{\citet}[1]{\cite{#1}}

\usepackage[utf8]{inputenc} 
\usepackage[T1]{fontenc}    
\usepackage{hyperref}       
\usepackage{url}            
\usepackage{booktabs}       
\usepackage{amsfonts}       
\usepackage{nicefrac}       
\usepackage{microtype}      
\usepackage[dvipsnames]{xcolor}         

\usepackage{amsmath,amssymb}
\usepackage{subcaption}

\usepackage{common}

\graphicspath{{assets/}}

\definecolor{orcidgreen}{HTML}{A6CE39}
\RequirePackage{tikz}
\usetikzlibrary{svg.path}
\def\orcidID#1{\smash{\href{http://orcid.org/#1}{\resizebox{1em}{1em}{\protect\begin{pgfpicture}
                                \pgfsetcolor{orcidgreen}
                                \pgfpathsvg{M106.301,35.435c0,19.572-15.863,35.436-35.436,35.436S35.43,55.007,35.43,35.435S51.293,0,70.865,0
                                        S106.301,15.863,106.301,35.435z}
                                \pgfusepath{fill}
                                \pgfsetcolor{white}
                                \pgfpathsvg{M59.321,32.39v13.398h-4.264V16.14h4.264V32.39z}
                                \pgfpathsvg{M65.578,16.112h11.572c9.772,0,15.724,7.226,15.724,14.839c0,7.004-4.816,14.838-15.779,14.838H65.578 V16.112z M69.841,41.941h6.562c8.305,0,12.098-5.039,12.098-10.99c0-3.654-2.215-10.99-11.877-10.99h-6.782V41.941z}
                                \pgfpathsvg{M57.189,54.758c-1.55,0-2.796-1.245-2.796-2.796c0-1.523,1.246-2.797,2.796-2.797s2.796,1.273,2.796,2.797 C59.985,53.485,58.74,54.758,57.189,54.758z}
                                \pgfusepath{fill}
\end{pgfpicture}}}}}

\title{Pantograph: A Machine-to-Machine Interaction Interface for Advanced Theorem Proving, High Level Reasoning, and Data Extraction in Lean~4}

\author{%
  Leni Aniva\orcidID{0000-0002-6033-9140} \quad Chuyue Sun\orcidID{0009-0005-9226-3688} \quad Brando Miranda\orcidID{0009-0008-5031-0126} \quad Clark Barrett\orcidID{0000-0002-9522-3084} \quad Sanmi Koyejo\orcidID{0000-0002-4023-419X} \\
  \texttt{\{aniva,chuyues,brando90,barrettc,sanmi\}@stanford.edu} \\
}
\institute{Stanford University}

\begin{document}

\maketitle

\begin{abstract}
	\emph{Machine-assisted theorem proving} refers to the process of conducting
structured reasoning to automatically generate proofs for mathematical theorems.
Recently, there has been a surge of interest in using machine learning models in
conjunction with proof assistants to perform this task. In this paper, we
introduce Pantograph, a tool that provides a versatile interface to the Lean~4
proof assistant and enables efficient proof search via powerful search
algorithms such as Monte Carlo Tree Search.  In addition, Pantograph enables
high-level reasoning
by enabling a more robust handling of Lean~4's inference steps.
We provide an overview of Pantograph's architecture and features.
We also report on an illustrative use case: using machine learning models
and proof sketches to prove Lean~4 theorems.
Pantograph’s innovative features pave the way for more advanced machine learning models to perform complex proof searches and high-level reasoning, 
equipping future researchers to design more versatile and powerful theorem provers.

\end{abstract}

\section{Introduction}
\label{sec:intro}
Proof assistants are used for a variety of tasks requiring strong
guarantees and rigorous reasoning.  High-profile applications include
formal verification of computer systems (e.g., sel4~\citep{sel4}) and
formalization of mathematics (e.g., \citep{hales2017formal}). Among proof
assistants, Lean~4 has recently accumulated significant momentum both among
mathematicians and non-mathematicians.  Its Mathlib library \citep{mathlib2019},
for example, is an extensive effort to formalize many branches of mathematics
and contains many non-trivial mathematical definitions and theorems.

A common challenge shared by all proof assistants is that completing proofs is
tedious and requires manual effort and expertise.  Machine learning offers one
potential avenue for addressing this challenge.  Indeed,  recent years have seen
several major efforts dedicated to using machine learning to automatically
search for proofs in proof assistants (e.g., \citep{wang2017}, \citep{yang2019},
\citep{crouse2020fol}, \citep{tactictoe2021}, \citep{neuroTactic2021},
\citep{hypertree2022}, \citep{dsp2022}, \citep{jiang2022thor},
\citep{wang2023dt}).  While these efforts have produced promising results, many
proofs are still beyond the reach of machine learning-based automation.

In order to continue to make progress in this area, several challenges need to
be addressed.  One of these challenges is the need for better interfaces
between proof assistants and machine learning systems.
%
%
\Comment{BM}{The novel contributions list is crucial for ML papers. I thus wrote in detail my suggestion + alternative ways to phrase things -- emphasising always motivation to why a feature is important. Importance has to be spelled out, especially to an ML crowd that knows very little about theorem proving TP.}
In this paper, we introduce
\textbf{Pantograph},\footnote{\url{https://github.com/stanford-centaur/PyPantograph}}
an API and Read-Eval-Print Loop (REPL) for Lean~4, whose primary goal is to
provide a convenient interface for training and evaluating theorem proving agents.
The name ``Pantograph'' alludes to the
process of recording a proof during proof search.\footnote{A Pantograph is a mechanism
for recording the movement of a pen while drawing in order to create a copy.}

The main motivation for creating Pantograph is to overcome the limitations of the
interface provided by the Lean~4 Language Server Protocol (LSP), which is the
standard interface provided for interactive use by a human user. Although the LSP
provides interactive feedback for a human operator of
Lean~4, it suffers from a number of problems as a machine interface. The LSP interface requires its
user to keep track of positions of a cursor in text, and a machine user would be
burdened with tracking these redundant data.
Moreover, there is no straightforward way to extract tactic training data from
the LSP interface or sketch out a proof to be finished by automation tactics.
In contrast, Pantograph is designed from the ground up as an efficient and
convenient interface for machine (and especially machine learning) agents.

The main contributions of Pantograph are:

\begin{enumerate}

\item Unlike prior work, the user can decide to solve goals independently. This
enables more powerful search algorithms such as Monte Carlo Tree Search (MCTS),
which have been successful in other domains (e.g., AlphaGo and AlphaZero
\cite{alphazero,alphago}), achieving superhuman performance on complex games
like Go, Chess, and Shogi.\footnote{\Rebuttal{A4}{Although these board games are not
equally difficult, the state-of-the-art algorithms for these board games all
involve MCTS.}} To do this, Pantograph handles metavariable coupling, which is a
phenomenon that complicates tree search~\citep{aesop}.
\item In contrast to prior work in Lean~4~\citep{leandojo}, Pantograph supports
the use of the advanced reasoning steps (called tactics) \lstinline{have},
\lstinline{let}, \lstinline{conv}, and \lstinline{calc}. These tactics are
crucial for supporting high-level  reasoning strategies like proof sketching
\citep{dsp2022}.
\item Pantograph fully supports essential data extraction tasks (e.g., it can
extract the before- and after-goal states of tactic executions, which are
usually not available in raw Lean~4 scripts). In addition, Pantograph
introduces several novel data extraction capabilities, including the ability to extract entire proof scripts with associated comments, which can be used for tasks like autoformalization, and the important ability to extract proof representations as programs, which allows for one-shot prediction of proofs.
\item Pantograph provides feedback from partially executed \lstinline{conv} and \lstinline{calc} tactics, which was not possible in preceding works.
\item Pantograph allows the user to resume an incomplete proof containing the
\lstinline{sorry} keyword in Lean~4. This is useful for machine learning models
which produce a proof draft before resolving the details in the proofs.
\item By making use of the novel features listed above, Pantograph can be used
to support the draft-sketch-proof (DSP) approach~\cite{dsp2022}.  An evaluation
of this approach on the important MiniF2F benchmark~\citep{minif2f} in Lean~4 is provided in Section~\ref{sec:eval}. To our
knowledge, this is the first implementation of DSP in Lean~4.
\end{enumerate}

As additional evidence of its usefulness, before this paper was even published,
research groups in both academia and industry were already
using Pantograph for machine-assisted theorem proving.

The rest of the paper is organized as follows. In Section~\ref{sec:background},
we cover background material on proof assistants and tree search.
We then discuss related work in Section~\ref{sec:related}.
Section~\ref{sec:features} gives an overview of the architecture and main features of Pantograph.
Section~\ref{sec:eval} illustrates and evaluates Pantograph’s capabilities
through an implementation of DSP in Lean~4.
Finally, Section~\ref{sec:conclusion} concludes.


\section{Background}
\label{sec:background}

\subsection{The Lean 4 Proof Assistant}
A \emph{proof assistant} is a computer program that can formulate and check
formal mathematical proofs. This includes Lean~4~\citep{lean4}, Coq \citep{coq},
Isabelle \citep{isabelle}, Aya \citep{aya}, and many others. \Rebuttal{A3}{These
programs operate by formulating mathematics as expressions and checking the
validity of the expressions via type-theoretic rules. Proof assistants may
differ in a number of ways, including
their syntax and the underlying variant of type theory they use.} In
the language of a proof assistant, every definition, theorem, or proof is a
value with a type. A value is represented by an \textbf{expression}.
\Comment{BM}{I don't think CIC needs to be mentioned.  Instead, my suggestion is
when explaining the background section is, what about the background helps the
(ML) reader understand the importance, impact and novelty of our work? the more
novel PL stuff we teach them the harder it will be for them to follow/remember.}
A proof of a theorem is a term whose type is the theorem.  A proof assistant
checks the validity of a proof of a theorem by evaluating its type and checking
that it matches the statement of the theorem.  It does this using a set of type
deduction rules.

For example, the commutativity of the logical OR ($\lor$) operation can be
written as the expression:
\begin{equation}
	\label{eqn:Or commutativity}
	\forall (p: \Prop), \forall (q: \Prop), \forall (h: p \lor q), q \lor p.
\end{equation}
This statement says that if $p$ and $q$ are Boolean propositions (of type $\Prop$
in Lean~4), then, given the hypothesis $h$ of type $p \lor q$, we can conclude
$q \lor p$.


\Comment{BM}{There are techincal terms like inhabitation and entailement, that I'm not sure if they are really needed. How are they helping the reader appreciate our contribution?}
\Comment{LA}{Fixed}
The notation in~\eqref{eqn:Or commutativity} is more verbose than what mathematicians
typically use. This is because proof assistants require the utmost unambiguity.
However, informally, the above expression could also be written using the more
concise notation:
\[ \forall\, p, q.\: p \lor q \to q \lor p
\]
A proof of~\eqref{eqn:Or commutativity} is an expression whose type is given
by~\eqref{eqn:Or commutativity}.  For example,
\begin{align*}
	\label{eqn:Proof of or commutativity}
	&\lambda (p, q: \Prop) (h: p \lor q) \\
	&\mapsto \lor.\operatorname{cases} h
		\,(\lambda h_p: p \mapsto \lor.\operatorname{inr} h_p)
		\,(\lambda h_q: q \mapsto \lor.\operatorname{inl} h_q)
\end{align*}
is a proof of the commutativity of OR.
\Comment{BM}{Yes! This is good to be said directly like this and should come way
earlier. Perhaps the first sentence. ML people have no idea about the
Curry-Howard correspondence and should be taught what this is and why we are
dealing with it.}
The type of a $\lambda$-expression is a $\forall$-expression. The $\lambda$'s in the
expression correspond to the three $\forall$'s in the statement of the
commutativity theorem. Intuitively, this expression says that when $p \lor q$ is
assumed to be true, proving $q \lor p$ requires proving $q \lor p$ when $p$ is
true and also when $q$ is true. This is signified by the special function
$\lor.\operatorname{cases}$, which is provided by Lean~4 as part of the support
for the $\lor$ operator. It represents the fact that deriving any value from
$p \lor q$ requires two functions, one to handle the case when $p$ is true, and
one to handle the case when $q$ is true.

Assuming $p$ is true, $\lor.\operatorname{inr}$ generates a proof of $q \lor p$
from a proof of the right operand $p$ ($\lor.\operatorname{inl}$ is similar but
requires a proof of the left operand $q$).

Expressions can also be constructed incrementally. For example, we could
postulate that the following expression has the type shown in Expression~\eqref{eqn:Or
commutativity}:
\[ \MVar{1} := \lambda (p:\Prop) \mapsto \MVar{2}[p]
\]
Then $\MVar{2}$ must have the type
\begin{equation}
	\label{eqn:Or commutativity step 1}
	\MVar{2} : \forall (q: \Prop), \forall (h: p \lor q), q \lor p \qquad
	\begin{cases}
		p: \Prop
	\end{cases}
\end{equation}
Here, $\MVar{1}$ and $\MVar{2}$ are \emph{metavariables}. A \textbf{metavariable} is a
variable, possibly unassigned, with a \textbf{context}. A \textbf{goal} (also
called a \textbf{hole}) is an
unassigned metavariable.    When writing proofs in Lean~4,
the \lstinline{sorry} keyword can be used as a placeholder for a hole.
A \textbf{free variable} in the \emph{context} of a
metavariable (e.g., the variable $p$ above) references a value assumed to be
true for this metavariable. $\MVar{2}$ is a goal. The \textbf{proof state}
consists of all metavariables, both those that are unassigned (i.e., the goals)
and those that are assigned.

Proof expressions, while easy for the proof assistant to check, are
difficult for a human operator to write. Thus, some proof
assistants such as Lean~4 also provide an alternative interface for theorem
proving, in which a proof can be executed via a series of \emph{tactics}. A
\textbf{tactic} changes the proof state by assigning an expression, possibly
containing new goals, to a goal in the current state.
In Lean~4, a tactic can transform one goal into a finite number of subgoals. A
tactic that generates no subgoals \emph{solves} the parent goal.  If all
subgoals produced by a tactic are solved, the goal is solved as well.
For example, suppose a variable $\MVar{1}$ has the type shown in
Expression~\eqref{eqn:Or commutativity}.  Executing the $\mathsf{intro}$ tactic
on $\MVar{1}$ results in the \emph{assignment}
$\MVar{1} := \lambda (p: \Prop) \mapsto \MVar{2}[p]$, where $\MVar{2}$ has the type in
Expression~\eqref{eqn:Or commutativity step 1}. $\MVar{2}$ becomes the new goal
that must be solved.



\Comment{BM}{I think it's important that we display visually the main concepts
they need to know about proof assistants e.g., see leandojo's pictures
\url{https://arxiv.org/pdf/2306.15626}, coqgym's pictures
\url{https://arxiv.org/pdf/1905.09381} or my (badly written) project report
\url{https://www.overleaf.com/read/xkvtgdqgwmqm\#68f5fb}}
\Comment{LA}{Is the diagram in the next section sufficient?}
\Comment{BM}{The background section should be self contained explanation of all the concepts an unexperienced reader in PL should know about Lean~4 to understand our paper.}

Some tactics can create interdependent metavariables. This is known as
\textbf{metavariable coupling} \citep{aesop}. For example, in order to prove
\[ \exists (x: \mathbb N), 2x + 5 \leq 10
\]
one would need to invoke the \lstinline{Exists.intro} lemma, which creates the
following goals in Lean~4:
\[
	\begin{aligned}
		\MVar{x}: &\mathbb N \\
		&2\MVar{x} + 5 \leq 10
	\end{aligned}
\]
where the second goal is now coupled to the first, since any solution of the
first goal will necessarily affect the second.

\subsection{Tree Search}

\textbf{Tree Search} refers to the process of searching through a tree,
each of whose nodes represents a potential solution to a problem, attempting to
find the best possible solution~\citep{mcts2012}. \textbf{Monte Carlo Tree Search}
(\textbf{MCTS}) is a class of tree search algorithms where in each iteration a
leaf node from the current search tree is selected and expanded. The selection
of this leaf node is driven by the \textbf{policy} of the tree search algorithm.

MCTS is used by AlphaGo \citep{alphago} and AlphaZero
\citep{alphazero} for playing board games and by HyperTree \citep{hypertree2022}
for proof search in Lean~3.

The tree structure that results from using tactics to prove theorems in Lean~4 is called an \emph{and-or tree} and contains two types
of nodes: \emph{goals} (Or), where solving at least one descendant suffices to solve
the goal, and \emph{goal states} (And) produced by tactics, where solving all
descendants is required. A Lean~4 proof begins with a single goal. A full proof
tree for the commutativity of OR is shown in Figure~\ref{fig:Proof tree for Or
commutativity}.
\Comment{BM}{I don't know if it's wise to foreshadow a important technical contribution like this because 1. the reader won't know what that is (In fact I'm not sure myself what that is), 2. they are wondering why this is mentioned without understanding why it's important. I think it's better to to have seperate sections explaining the importance of each contribution. A. One for proof assistance in general and thus proof terms B. One for coupling and it's importance C. high level reasoning that have enables. I think the background section should set up the reader to understand our 3 main contributions. }
\Comment{LA}{Fixed}
When applying Monte Carlo Tree Search two theorem proving, two functions are required:
the \textbf{policy function} decides which node (i.e., goal) to explore next,
and the \textbf{tactic function} decides which tactic to use on that goal.

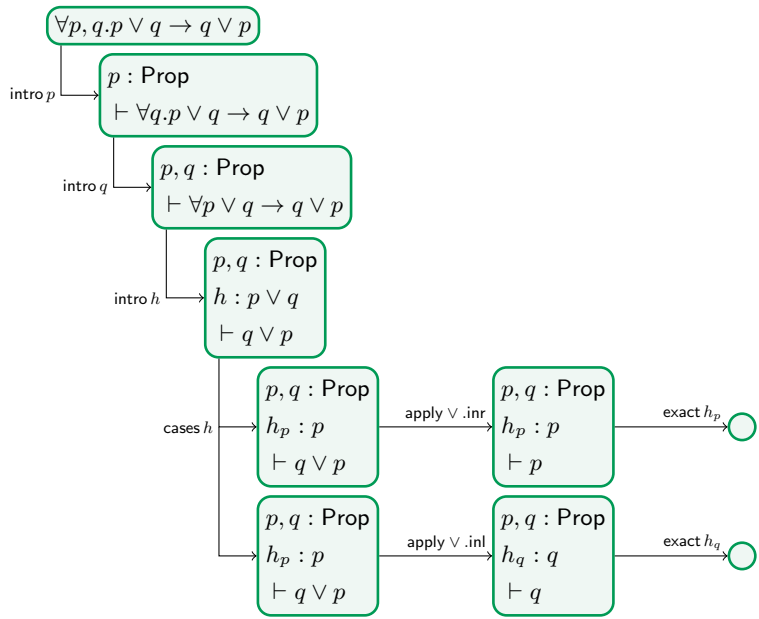
\begin{figure}
	\centering
	\begin{tikzpicture}[
			xscale=1,
			Anchored/.style={below right},
			Tactic/.style={scale=0.7}]
		\coordinate (arrow-shift) at (0.2,0);
		\coordinate (child-shift) at (0.5,-.1);

		\node[Goal,Anchored] (g0) at (0,0) {$\forall p,q. p \lor q \to q \lor p $};
		\coordinate (parent) at ($(g0.south west) + (arrow-shift)$);
		\node[Goal,Anchored] (g1) at ($(parent)+(child-shift)$) {$\begin{aligned}&p: \Prop\\&\vdash \forall q. p \lor q \to q \lor p\end{aligned}$};
		\draw[->] (parent) -- (parent |- g1.west)
			node[Tactic,left] {$\mathsf{intro}\,p$}
			-- (g1.west);

		\coordinate (parent) at ($(g1.south west) + (arrow-shift)$);
		\node[Goal,Anchored] (g2) at ($(parent) + (child-shift)$)
			{$\begin{aligned}&p,q: \Prop\\&\vdash \forall p \lor q \to q \lor p\end{aligned}$};
		\draw[->] (parent) -- (parent |- g2.west)
			node[Tactic,left] {$\mathsf{intro}\,q$}
			-- (g2.west);

		\coordinate (parent) at ($(g2.south west) + (arrow-shift)$);
		\node[Goal,Anchored] (g3) at ($(parent)+(child-shift)$)
			{$\begin{aligned}&p,q: \Prop\\&h: p \lor q\\&\vdash q \lor p\end{aligned}$};
		\draw[->] (parent) -- (parent |- g3.west)
			node[Tactic,left] {$\mathsf{intro}\,h$}
			-- (g3.west);

		\coordinate (parent) at ($(g3.south west) + (arrow-shift)$);
		\node[Goal,Anchored] (g4a) at ($(parent)+(child-shift)$)
			{$\begin{aligned}&p,q: \Prop\\&h_p: p\\&\vdash q \lor p\end{aligned}$};
		\node[Goal,Anchored] (g4b) at ($(g4a.south west)+(0,-0.1)$)
			{$\begin{aligned}&p,q: \Prop\\&h_p: p\\&\vdash q \lor p\end{aligned}$};
		\draw[->] (parent) -- (parent |- g4a.west)
			node[Tactic,left] {$\mathsf{cases}\,h$}
			-- (g4a.west);
		\draw[->] (parent |- g4a.west) -- (parent |- g4b.west) -- (g4b.west);

		\coordinate (parent) at (g4a.east);
		\node[Goal,right] (g5a) at ($(parent)+(1.5,0)$)
			{$\begin{aligned}&p,q: \Prop\\&h_p: p\\&\vdash p\end{aligned}$};
		\draw[->] (parent) -- (g5a.west)
			node[Tactic,above left] {$\mathsf{apply \lor.\mathsf{inr}}$};

		\coordinate (parent) at (g4b.east);
		\node[Goal,right] (g5b) at ($(parent)+(1.5,0)$)
			{$\begin{aligned}&p,q: \Prop\\&h_q: q\\&\vdash q\end{aligned}$};
		\draw[->] (parent) -- (g5b.west)
			node[Tactic,above left] {$\mathsf{apply \lor.\mathsf{inl}}$};

		\coordinate (parent) at (g5a.east);
		\node[Goal,right] (g6a) at ($(parent)+(1.5,0)$) {};
		\draw[->] (parent) -- (g6a.west)
			node[Tactic,above left] {$\mathsf{exact}\,h_p$};

		\coordinate (parent) at (g5b.east);
		\node[Goal,right] (g6b) at ($(parent)+(1.5,0)$) {};
		\draw[->] (parent) -- (g6b.west)
			node[Tactic,above left] {$\mathsf{exact}\,h_q$};
	\end{tikzpicture}
	\caption{A proof tree for Expression~\eqref{eqn:Or commutativity}}
	\label{fig:Proof tree for Or commutativity}
\end{figure}

The main motivation for creating Pantograph is to create an interface that can
easily be used by machine learning systems aiming to exploit this incremental
tree structure to conduct mathematical reasoning. \Rebuttal{C4}{Potential
applications of Pantograph include automatic verified program generation,
rigorous reasoning for language models, and autoformalization of mathematics
results.}

\section{Related Work}
\label{sec:related}


The closest related work is \textbf{LeanDojo}~\citep{leandojo}, which provides a Python interface for machine interaction
with Lean~4.  A user can use this interface to execute tactics on a current
proof state in a Lean~4 process.
LeanDojo can be used to save and resume proof states.
LeanDojo can
also run commands such as \lstinline{#eval} and \lstinline{#check} and extract
goal-tactic pairs over an existing proof.

Pantograph has several key architectural improvements over LeanDojo. First of
all, it is written entirely in Lean~4.  This removes the need for external
dependencies (such as Docker) and also improves the speed of
interaction.
Also, some tactics not supported by LeanDojo, such as \lstinline{have},
are available in Pantograph. Other tactics, such as \lstinline{conv} and \lstinline{calc}, can only
be used monolithically to solve goals in LeanDojo, whereas Pantograph supports
incremental exploration using these tactics, allowing a user to obtain
feedback at each step, even if a goal is not solved.
Moreover, Pantograph efficiently handles the problem of metavariable coupling
(see Section~\ref{sec:metavar}),
empowering ML models to work on interdependent proof branches without risking
inconsistency.
\Comment{BM}{instead of "which will be discussed in the next section" provide a 1 sentence sumary: "which empowers ML models to search different yet depedent proof branches indepdently without risking inconsistent proof creation".}
\Comment{LA}{Fixed}

Like LeanDojo, Pantograph can extract information from existing proofs,
including training data based on triples of goal states, tactics, and
post-tactic goal states.
It can also extract comment data and arbitirary expressions, both of which may
be useful as additional training data.
\Comment{BM}{I think it's better to say "we can do all lean dojo can do in data extraction" + "we can also extract lean scripts with comment useful for AF and proof terms for additional trianing data".}
\Comment{LA}{Fixed}
However, LeanDojo's data extraction and proof execution units are essentially
separate, which makes it impossible to extract an incomplete proof and resume
from it, whereas Pantograph supports this use case.

In~\citep{hypertree2022}, Lample et al. implement the aforementioned
and-or tree search structure to solve goals in Lean~3. In this work, the relation
between goals and tactics is a \emph{hypertree}. This enables efficient proof
search via a variant of Monte Carlo Tree Search. The policy and tactic functions
are both provided by Large Language Models (LLMs). Pantograph is compatible
with this approach, as it gives the user control over both policy and tactic functions.

Draft-Sketch-Prove (DSP) \citep{dsp2022} is a \emph{neural theorem prover} which uses
an approach
based on \emph{drafting}. Instead of directly generating Lean~4 tactics, the neural
theorem prover generates intermediate goals in an informal language (draft).
Then it translates these goals into Isabelle (sketch), and finally a
\emph{hammer} tactic from Isabelle solves the goals (prove). Pantograph's drafting
feature supports this technique as well, allowing the Draft-Sketch-Prove
algorithm to be implemented in Lean~4 (see Section~\ref{sec:dsp}).

Aesop \citep{aesop} is a proof automation (\emph{hammer}) tactic based on
tree-search. Aesop is not based on machine learning and takes metavariable
coupling into account when solving goals. When a goal gets solved in Aesop, all
of the goals coupled to this goal are brought back into scope. This is known
as \emph{copying}. Pantograph's approach to metavariable coupling is based on
Aesop's technique, but extends it by allowing the user to determine which
metavariable to solve next.

CoqGym \citep{yang2019} is similar to LeanDojo but for the Coq theorem prover
instead of Lean~4. CoqGym stores proofs in a tree structure and allows the user
agent to execute tactics.  Optionally, CoqGym can serialize proof terms into
S-expressions. These same features are supported by Pantograph, but for Lean~4.
\Rebuttal{B3}{Pantograph also has some features unsupported by CoqGym, such as
the ability to implement draft-sketch-prove and to handle metavariable coupling.}

\section{Architecture and Features}
\label{sec:features}

\begin{figure}[t]
	\centering
	\begin{tikzpicture}[
			xscale=3, yscale=1.2,
			Label/.style={scale=0.7},
		]
		\filldraw[fill=CadetBlue!30,dashed,thin,rounded corners]
			(-1.5,-0.3) rectangle (1.5,1.3);
		\node[draw,rounded rectangle] (kernel) at (0,0) {Lean Kernel};
		\node[draw,rectangle] (project) at (0,-1) {Lean Project};
		\draw[->,dashed] (kernel) -- (project)
			node[midway,right,Label] {Reads};

		\node[draw,rounded rectangle] (p) at (-1,0) {Pantograph};
		\node[draw,rounded rectangle] (repl) at (-1,1) {REPL};
		\draw[->] (p) -- (kernel);
		\draw[->] (repl) -- (p);

		\node[draw,rounded rectangle] (lsp) at (1,0) {Lean LSP};
		\node[draw,rounded rectangle] (ide) at (1,1) {IDE};
		\draw[->] (lsp) -- (kernel);
		\draw[->] (ide) -- (lsp);
		\draw[->] (p) to[bend left=30] (lsp);

		\node (ml) at (-1,2) {User (e.g., Neural Network)};
		\node (user) at (1,2) {Human Operator};
		\draw[->] (ml) -- (repl);
		\draw[->] (user) -- (ide);
	\end{tikzpicture}
	\caption{System architecture of Pantograph. A solid arrow indicates
          that the component at the arrow source calls functions in the
          component that is the arrow's target. A human operator interacts
	with Lean~4's kernel via the IDE, but a machine learning agent can
        interact via one of Pantograph's interfaces.}
	\label{fig:System architecture of Pantograph}
\end{figure}
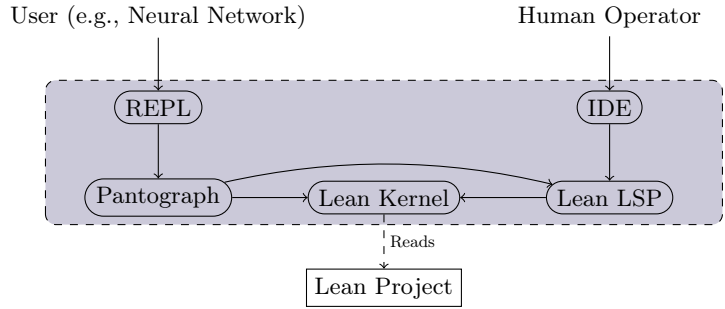
\Comment{BM}{I personally feel figs drawn by latex aren't usually high quality.
I recommend a tood like
\url{https://excalidraw.com/\#json=uvFvrjLVUdXppNmH-g5_j,3PBV0-2Z--vUTHC07OaSgg} to draw
the system design or see my PyCoq project for how I explain proof terms
\url{https://www.overleaf.com/read/xkvtgdqgwmqm\#68f5fb}}
\Comment{BM}{This is more of a brain storming idea, but I feel the ML user might not care so much how internally we talk to Lean~4 but instead the API and thus the increased capabilities their models have. I think that is how CoqGym and LeanDojo present their work. What so you think? Otherwise I feel the ML reader is left think "ok of course, its an RL env for lean, of course it has to talk with lean, why is this being said to me, isn't the fact it talks to Lean~4 a give?". }
\Comment{LA}{Obsolete since we are not submitting to an ML conference}

\begin{figure}
	\centering
	\begin{tikzpicture}[
			Function/.style={draw,rectangle,below right},
			Monad/.style={black!50,scale=0.8},
		]
		\newcommand{\Callee}[2]{
			\coordinate (pos) at ($(pos)+(shift)$);
			\node[Function] (#1) at (pos) {#2};
		}
		\newcommand{\LeanCallee}[2]{
			\coordinate (pos) at ($(pos)+(shift)$);
			\node[Function,fill=ForestGreen!20] (#1) at (pos) {#2};
		}
		\newcommand{\Call}[2]{%
			\coordinate (src) at ($(#1.south west)+(0.1,0)$);
			\draw[->] (src) -- (src |- #2.west) -- (#2);
		}%

		\coordinate (pos) at (0,0);
		\coordinate (shift) at (0.2,-.5);
		\node[Function] (f0) at (pos) {\lstinline{main}};
		\node[Monad,right] at (f0.east) {\lstinline{IO}};
		\Callee{f1}{\lstinline{loop}}
		\node[Monad,right] at (f1.east) {\lstinline{CoreM}};
		\Call{f0}{f1}
		\Callee{f2}{\lstinline{Pantograph.execute}}
		\node[Monad,right] at (f2.east) {\lstinline{CoreM}};
		\Call{f1}{f2}
		\Callee{f3}{\lstinline{Pantograph.GoalState.tryTactic}}
		\node[Monad,right] at (f3.east) {\lstinline{TermElabM}};
		\Call{f2}{f3}
		\LeanCallee{f4}{\lstinline{Lean.Elab.Tactic.evalTactic}}
		\node[Monad,right] at (f4.east) {\lstinline{TacticM}};
		\Call{f3}{f4}
	\end{tikzpicture}
	\caption{Call hierarchy in Pantograph during the execution of a normal
	tactic. The text on the right indicates the Lean~4 monad each function runs
	in.}
	\label{fig:Call hierarchy during normal tactic}
\end{figure}
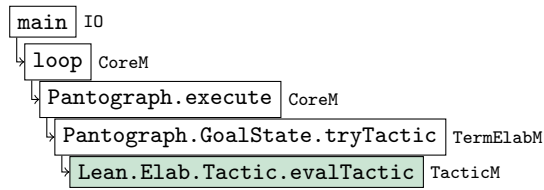

Pantograph is implemented entirely in Lean~4 with no external dependencies.
Figure~\ref{fig:System architecture of Pantograph} shows an overview of the
operation of Pantograph. The user, which is often a machine learning model,
calls Pantograph's functions via one of its interfaces.
Pantograph provides three interfaces:
%
($i$) a Python interface called PyPantograph;
($ii$) a REPL via the \textsf{pantograph-repl} executable; and
($iii$) a library via the C Foreign Function Interface (FFI).
When the user executes a tactic, Pantograph calls the Lean~4 kernel's
\lstinline{Elab.Tactic.evalTactic} function. Internally, many of Lean~4's
functions are \emph{monads}, which are abstract structures enabling state
manipulation in an otherwise functional language. Lean~4's monad hierarchy (from
the order of most to least general) has the order \lstinline{IO},
\lstinline{CoreM}, \lstinline{MetaM}, \lstinline{Elab.TermElabM}, and
\lstinline{Elab.Tactic.TacticM}.
Figure~\ref{fig:Call hierarchy during normal tactic} outlines the most
important functions called during the execution of a tactic via Pantograph.
\Comment{BM}{Sadly, I don't remember what a monad is. If I don't remember and I use Lean~4/TP like stuff often, an ML reader won't know either, and will be thrown off by the term being said casually.}

Other features of Pantograph call into the Lean~4 Language Server Protocol
(LSP), the Lean~4 parser, and the Lean~4 compiler.
In particular, Pantograph intercepts the Lean~4 compiler state when it processes
Lean~4 source code, enabling it to extract information that is otherwise only available
via the IDE.

In the rest of this section, we provide details about the features available in
Pantograph.  We discuss Pantograph's support for the following features:
(\textit{i}) both expression-based and tactic-based proof;
(\textit{ii}) tree search;
(\textit{iii}) custom handling of metavariable coupling;
(\textit{iv}) the extraction of tactic training data; and
(\textit{v}) drafting.

\subsection{Expressions and Tactics}

Pantograph enables AI agents to use the same tactics as a human operator can
while interacting with Lean.  Human-written proofs in Lean~4 (e.g., in Mathlib)
are often a mixture between expressions and tactics. As mentioned above, tactics
are used to reduce a goal to one or more new subgoals. To see how expressions
can be used, consider the following example.
\begin{lstlisting}
example : exists x, x + 2 = 8 := by
  let a : Nat := 3 * 2
  exists a
\end{lstlisting}
In this example, the required witness for $x$ is directly constructed as the expression $3 \cdot 2$.
Pantograph supports seamlessly switching between expression-based and
tactic-based proof by providing a custom
\lstinline{expr} tactic. This tactic takes an expression
$e[\MVar{1},\MVar{2},\dots]$ and assigns it to the current goal. The holes
$\MVar{1},\MVar{2},\dots$ then become the new goals.

There are 3 views of a proof:
\begin{enumerate}
\item \textbf{Presentation View}: A proof written for presentation and
	verification. Coupling does not exist. It may contain values with puzzling
	origins such as complex bounds that are not apparent before reading the entire
proof.
\item \textbf{Search View}: A proof viewed as the trajectory of a proof search
	agent traverses through while finding the proof. It may contain backtracking,
	coupling, and goal selection.
\item \textbf{Kernel View}: A proof viewed as a set of metavariables.
\end{enumerate}
Pantograph enables agents to operate in the search view and handles proofs
internally in the kernel view.


Lean~4 includes sophisticated tactics, like \lstinline{conv} and
\lstinline{calc}, which are \emph{composite} in the sense that they  are used to
compose sequences of other tactics.  While these tactics can be executed
monolithically by supplying the full sequence of tactics to be composed, human
operators of Lean~4 often rely on Lean~4's interactive interface to
\emph{incrementally} explore possible sequences, obtaining feedback at each
step.  Pantograph provides a custom tactic called \lstinline{goal.tactic}, which
can partially execute a \lstinline{conv} or \lstinline{calc} tactic and provide
feedback from this partial execution.
As an example, consider the following use of the \lstinline{calc} tactic, which
is used in Lean~4 to compose a series of transitivity steps.
\begin{lstlisting}
example (a b c : Nat) : a + b = b + c := by
  calc a + b = a + a := sorry
    _ = b + b := sorry
  sorry
\end{lstlisting}
Here, the goal \lstinline{a + b = b + c} is not provable by \lstinline{calc},
but a user can still partially execute the tactic by applying just the first
line and seeing what the result is.  In this case, the result of executing just
the first line results in the following new goal.
\begin{lstlisting}
a b c : Nat
|- a + a = b + c
\end{lstlisting}
Pantograph supports this partial execution model and can return the new goal
shown above.

Pantograph also supports the \lstinline{have} and \lstinline{let} tactics.
These tactics define temporary expressions in a local scope and are
indispensable when developing proofs by hand.
For example, consider the following snippet.
\begin{lstlisting}
example (n: Nat), n + 0 = 0 + n := by
	have h1 : n + 0 = n := sorry
	sorry
\end{lstlisting}
The use of \lstinline{have} introduces a new expression and a new goal.  The
two \lstinline{sorry} expressions create two holes corresponding to the two goals shown below.
\begin{lstlisting}
n : Nat
|- n + 0 = n
n : Nat
h1 : n + 0 = n
|- n + 0 = 0 + n
\end{lstlisting}
The Pantograph repository contains documentation and examples for these tactics.


In order to be friendly towards searching methods such as Monte Carlo Tree
Search \citep{mcts2012}, Pantograph provides an interface for incrementally
executing tactics. If a tactic creates more than one goal, it is called a
\emph{branching} tactic. 
When more than one goal exists in a proof state, Pantograph provides the option to
choose which goal to apply a tactic to.

\Rebuttal{B1}{If a tactic cannot execute for some reason, Pantograph outputs an
error message corresponding to what a human operator would see during interaction
with Lean's LSP.}

\subsection{Tree Search}

As mentioned above, tree search is a common search technique and is utilized in
various proof search approaches such as HyperTree
\citep{hypertree2022} and Aesop \citep{aesop}. Since each tactic produces zero
or more goals, the search structure of applying tactics to goals can be viewed
as an And-Or tree (in the absence of metavariable coupling, see Section~\ref{sec:metavar}). When the current
proof state has multiple goals, Pantograph allows the user to choose
which goal to attempt next, i.e., it allows user-defined policy functions.

This naturally leads to the question of the fate of sibling goals. Suppose there
are two goals $[\MVar{1}, \MVar{2}]$ in the current proof state, and the user
applies a tactic to $\MVar{1}$, generating $\MVar{3}$. The status of $\MVar{2}$
depends on the \emph{automatic mode} option. Automatic mode is turned on by
default, which means sibling goals are carried forward to the next proof state.
Hence, with automatic mode on, the next proof state would contain
$[\MVar{3}, \MVar{2}]$, with all goals present and active. If the user disables
automatic mode, the proof state instead becomes $[\MVar{3}]$. The goal
$\MVar{2}$ becomes \emph{dormant}. Dormant goals are unassigned metavariables
that do not appear in the current proof state. Note that dormant goals are an
artifact of Pantograph's manual tree search capability: they do not occur when
using Lean~4 through the interactive interface.
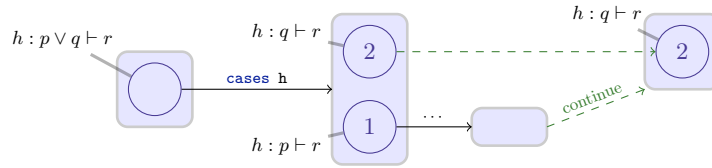
\begin{figure}
	\centering
	\begin{tikzpicture}[
			State/.style={line width=1pt,blue!10,rounded corners, draw=black!20},
			Goal/.style={draw,circle,minimum size=7mm,BlueViolet},
			TacticText/.style={midway,above,scale=0.7},
			GoalText/.style={scale=0.8},
			textRel/.style={very thick,opacity=0.3},
			tactic/.style={->},
			resume/.style={OliveGreen,dashed,->},
			ResumeText/.style={midway,above,sloped,scale=0.7},
			coupling/.style={Fuchsia,->},
		]
		\coordinate (textrel) at (-1,.5);

		\coordinate (state) at (0,0);
		\coordinate (anchor) at ($(state)+(0.5,0)$);
		\filldraw[State] ($(state) - (0,0.5)$) rectangle ++(1,1);
		\node[Goal] (g0) at (anchor) {};
		\node[GoalText,above left=.3] (g0t) at (g0.north west)
			{$h : p \lor q \vdash r$};
		\draw[textRel] (g0) -- (g0t);

		\coordinate (state) at ($(g0.east) + (2,0)$);
		\coordinate (anchor) at ($(state)+(0.5,0)$);
		\filldraw[State] ($(state) - (0,1)$) rectangle ++(1,2);
		\node[Goal] (g10) at ($(anchor) + (0,-.5)$) {\MVar{1}};
		\node[GoalText,left=.3] (g10t) at (g10.south west) {$h : p \vdash r$};
		\draw[textRel] (g10) -- (g10t);
		\node[Goal] (g11) at ($(anchor) + (0, .5)$) {\MVar{2}};
		\node[GoalText,left=.3] (g11t) at (g11.north west) {$h : q \vdash r$};
		\draw[textRel] (g11) -- (g11t);
		\draw[tactic] (g0) -- (state)
			node[TacticText] {\lstinline{cases h}};

		\coordinate (state) at ($(g10.east)+(1,0)$);
		\filldraw[State] ($(state) - (0,0.25)$) rectangle ++(1,.5);
		\draw[tactic] (g10) -- (state)
			node[TacticText] {$\cdots$};
		\coordinate (stateFin) at ($(state)+(1,0)$);

		\coordinate (state) at ($(g11.east)+(3.3,0)$);
		\coordinate (anchor) at ($(state)+(0.5,0)$);
		\filldraw[State] ($(state) - (0,.5)$) rectangle ++(1,1);
		\node[Goal] (g2) at (anchor) {\MVar{2}};
		\node[GoalText] (g2t) at ($(g2) + (textrel)$) {$h : q \vdash r$};
		\draw[textRel] (g2) -- (g2t);

		\draw[resume] (stateFin) -- ($(state)+(0,-.5)$) node[ResumeText] {continue};
		\draw[resume] (g11) -- (g2);
	\end{tikzpicture}
	\caption[2 becomes dormant after a tactic is applied to its sibling]{$\MVar{2}$ becomes dormant after a tactic is applied to $\MVar{1}$.
	It must be brought back into scope with \lstinline{goal.continue} before the
proof can finish. The ellipses ($\ldots$) are plalceholders for some combination
		of tactics which eventually solves the descendant of $\MVar{1}$.}
	\label{fig:Continuation}
\end{figure}
Dormant goals must either be tracked by the user or brought back into the proof
state using the \lstinline{goal.continue} command, as shown in
Figure~\ref{fig:Continuation}.

To summarize, in \emph{automatic mode}, goals are immediately continued after a tactic
execution. Goals will never become dormant in automatic mode. This provides a
gym-like environment to its user. Users who wish to handle tree search manually
should disable this mode.

\subsection{Metavariable Coupling}
\label{sec:metavar}


Recall that a proof state may contain 0 or more goals, and 
metavariable coupling~\citep{aesop} refers to inter-dependencies between goals
in a proof state.
Metavariable coupling arises naturally in many contexts. For
example, applying the transitivity axiom of $\leq_{\mathbb N}$ to the goal $2 \leq 5$
results in the following goals.
\begin{align*}
	\MVar{1} &: 2 \leq \MVar{z} \\
	\MVar{2} &: \MVar{z} \leq 5 \\
	\MVar{z} &: \mathbb N
\end{align*}
Because $\MVar{z}$ appears in all three goals, these goals are all coupled.
This complicates proof search because if an assignment is made to $z$ in one
goal, it will propagate to all of the other coupled goals. In this case,
the other two goals will no longer be coupled, but they will contain the
assignment made to $z$.

Pantograph provides explicit information about which goals are coupled.
Since there are multiple possible ways of
handling coupling, the choice of what to do with the coupling is left to the
user. One method employed by \citep{aesop} is \emph{copying}, where
coupled goals are solved sequentially to avoid conflicts.

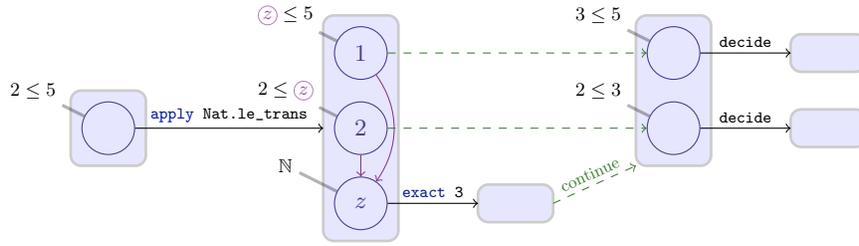
\begin{figure}
	\centering
	\begin{tikzpicture}[
			State/.style={line width=1pt,blue!10,rounded corners, draw=black!20},
			Goal/.style={draw,circle,minimum size=7mm,BlueViolet},
			TacticText/.style={midway,above,scale=0.7},
			GoalText/.style={scale=0.8},
			textRel/.style={very thick,opacity=0.3},
			tactic/.style={->},
			resume/.style={OliveGreen,dashed,->},
			ResumeText/.style={midway,above,sloped,scale=0.7},
			coupling/.style={Fuchsia,->},
		]
		\coordinate (textrel) at (-1,.5);

		\coordinate (state) at (0,0);
		\coordinate (anchor) at ($(state)+(0.5,0)$);
		\filldraw[State] ($(state) - (0,0.5)$) rectangle ++(1,1);
		\node[Goal] (g0) at (anchor) {};
		\node[GoalText] (g0t) at ($(g0) + (textrel)$) {$2 \leq 5$};
		\draw[textRel] (g0) -- (g0t);

		\coordinate (state) at ($(g0.east) + (2.5,0)$);
		\coordinate (anchor) at ($(state)+(0.5,0)$);
		\filldraw[State] ($(state) - (0,1.5)$) rectangle ++(1,3);
		\node[Goal] (g10) at ($(anchor) + (0,-1)$) {$\MVar{z}$};
		\node[GoalText] (g10t) at ($(g10) + (textrel)$) {$\mathbb N$};
		\draw[textRel] (g10) -- (g10t);
		\node[Goal] (g11) at ($(anchor) + (0, 0)$) {$\MVar{2}$};
		\node[GoalText] (g11t) at ($(g11) + (textrel)$) {$2 \leq \MVar{z}$};
		\draw[textRel] (g11) -- (g11t);
		\node[Goal] (g12) at ($(anchor) + (0, 1)$) {$\MVar{1}$};
		\node[GoalText] (g12t) at ($(g12) + (textrel)$) {$\MVar{z} \leq 5$};
		\draw[textRel] (g12) -- (g12t);
		\draw[coupling] (g11) -- (g10);
		\draw[coupling] (g12) edge[bend left=35] (g10);
		\draw[tactic] (g0) -- (state)
			node[TacticText] {\lstinline{apply Nat.le_trans}};

		\coordinate (state) at ($(g10.east)+(1.2,0)$);
		\filldraw[State] ($(state) - (0,0.25)$) rectangle ++(1,.5);
		\draw[tactic] (g10) -- (state)
			node[TacticText] {\lstinline{exact 3}};
		\coordinate (stateFin) at ($(state)+(1,0)$);

		\coordinate (state) at ($(g11.east)!.5!(g12.east)+(3.3,0)$);
		\coordinate (anchor) at ($(state)+(0.5,-.5)$);
		\filldraw[State] ($(state) - (0,1)$) rectangle ++(1,2);
		\node[Goal] (g21) at (anchor) {};
		\node[GoalText] (g21t) at ($(g21) + (textrel)$) {$2 \leq 3$};
		\draw[textRel] (g21) -- (g21t);
		\node[Goal] (g22) at ($(anchor)+(0,1)$) {};
		\node[GoalText] (g22t) at ($(g22) + (textrel)$) {$3 \leq 5$};
		\draw[textRel] (g22) -- (g22t);

		\draw[resume] (stateFin) -- ($(state)+(0,-1)$) node[ResumeText] {continue};
		\draw[resume] (g11) -- (g21);
		\draw[resume] (g12) -- (g22);

		\coordinate (state) at ($(g21.east)+(1.2,0)$);
		\filldraw[State] ($(state) - (0,0.25)$) rectangle ++(1,.5);
		\draw[tactic] (g21) -- (state)
			node[TacticText] {\lstinline{decide}};
		\coordinate (state) at ($(g22.east)+(1.2,0)$);
		\filldraw[State] ($(state) - (0,0.25)$) rectangle ++(1,.5);
		\draw[tactic] (g22) -- (state)
			node[TacticText] {\lstinline{decide}};
	\end{tikzpicture}

	\caption{In this diagram, rectangular boxes are proof states, and circles are
goals. Each proof state has 0 or more goals. A state with no goals is considered solved. If
all descendant goals of a state become solved, the state itself becomes solved.}
	\label{fig:Proving 2 le 5}
\end{figure}
Figure~\ref{fig:Proving 2 le 5} gives a full example of the above proof,
conducted with automatic mode \emph{off}.
The application of the transitivity tactic creates a proof state with three
goals.  Using the \lstinline{exact 3} tactic on the $\MVar{z}$ goal results in
a solved proof state.  Applying \lstinline{goal.continue} then brings goals
$\MVar{1}$ and $\MVar{2}$ back into the proof state, where they are no longer
coupled.  Each can be discharged with an additional tactic such as \lstinline{decide}.

\subsection{The Environment}

All running instances of Lean~4, including instances running behind the
LSP or the Pantograph front end, maintain a library of active
symbols known as the \emph{environment}. Internally, Lean~4 stores all theorem
statements and proofs in the environment as expressions, regardless of how they
were constructed.  The user can extract the proof of any theorem in the current
environment via \lstinline{env.inspect}, which will produce an expression
similar to Expression~\eqref{eqn:Or commutativity}.

After a proof concludes, the user can extract the proof expression of the root
proposition using the \lstinline{goal.print} command. These expressions can then
be inserted back into the current environment using the \lstinline{env.add}
command. Adding a lemma to the environment makes it accessible as a step in future proofs. Note
that adding lemmas to the environment cannot be done while in the middle of an
incomplete proof.

Like CoqGym \citep{yang2019}, Pantograph can optionally output proof expressions
in S-expression format by turning on the \lstinline{printExprAST} option via
\lstinline{options.set}. The user can also change Lean~4's expression
pretty-printing options by providing command line parameters to the Pantograph REPL.
For example, to turn off all pretty-printing, use \lstinline{--pp.all=true}.

\subsection{Tactic Training Data}

The Lean~4 community has produced several large collections of theorems with
human-written formal proofs, e.g., Mathlib \citep{mathlib2019}. These collections can be used to train theorem
proving agents. The \lstinline{frontend.process} command runs the Lean~4
compiler on a Lean~4 file, collects all tactics in the file, and returns
them as a list of (before, after, tactic) triplets. These triplets are conveniently
presented in a format conducive to offline reinforcement learning training.
Pantograph also outputs information about the starting and ending positions (in
the file) of each Lean~4 command in case the user is interested in processing
comments or other metadata.
Below is an example of one extracted tactic triple.
\begin{lstlisting}
{
	"goalBefore":"(*$\vdash \forall (p, q : \Prop), p \lor q \to q \lor p$ *)",
	"goalAfter":"(* $p: \Prop$ *)\n(* $\vdash \forall (q : \Prop), p \lor q \to q \lor p$ *)",
	"tactic":"intro p"
}
\end{lstlisting}

\subsection{Drafting}
\label{sec:dsp}

\textbf{Drafting} refers to a theorem proving technique which starts by
generating a proof outline, instead of building a full
proof step by step.  A \emph{draft} proof first consists of an overview with
holes.  Draft proofs are resolved by proving the
individual goals corresponding to the holes in the proof.
For example, consider the task of proving the commutativity of
addition in Peano arithmetic. One approach would be to write a proof based on
induction, using the inductive hypothesis $n + m = m + n$ to prove the
inductive step $m + (n + 1) = (m + n) + 1$. As stated, this proof is not rigorous or
detailed enough for Lean~4, but it \emph{can} be written as
a draft proof:
\begin{lstlisting}
theorem add_comm : forall n m : Nat, n + m = m + n := by
   intros n m
   induction n with
   | zero =>
     have h_base: 0 + m = m := sorry
     have h_symm: m + 0 = m := sorry
     sorry
   | succ n ih =>
     have h_inductive: n + m = m + n := sorry
     have h_pull_succ_out_from_right: m + Nat.succ n = Nat.succ (m + n) := sorry
     sorry
\end{lstlisting}
The placeholders for intermediate goals have marked with the
\lstinline{sorry} keyword.

Pantograph supports drafting in two ways. The first is via the \lstinline{have}
tactic. This tactic introduces a lemma or intermediate claim and creates a
new goal corresponding to the lemma.


The other way Pantograph supports drafting is via \lstinline{sorry}-extraction.
Pantograph can find all occurrences of \lstinline{sorry} in a proof or definition and convert
them to goals. For example, when \lstinline{add_comm} from the above proof is
fed into Pantograph's \lstinline{frontend.process} command, it generates the following list
of goals:
\begin{lstlisting}
m : Nat
|- 0 + m = m
m : Nat
h_base : 0 + m = m
|- m + 0 = m
m : Nat
h_base : 0 + m = m
h_symm : m + 0 = m
|- 0 + m = m + 0
m : Nat
n : Nat
ih : n + m = m + n
|- n + m = m + n
m : Nat
n : Nat
ih : n + m = m + n
h_inductive : n + m = m + n
|- m + n.succ = (m + n).succ
m : Nat
n : Nat
ih : n + m = m + n
h_inductive : n + m = m + n
h_pull_succ_out_from_right : m + n.succ = (m + n).succ
|- n + 1 + m = m + (n + 1)
\end{lstlisting}
The user can then execute tactics on these goals as
they see fit.
This feature is appealing for machine learning agents, since it allows an
agent (e.g., a Generative AI agent like an LLM) to effectively draft the next
step of the proof without having to dive into details about its execution.
\Rebuttal{B2}{If a sketch contains type errors that cannot be rectified,
Pantograph forwards the error generated by the Lean~4 Kernel
to the user.}

\subsection{Limitations}

\Rebuttal{C1}{Pantograph is limited by the functionalities available in Lean.
For example, if a tactic has a bug and discards a metavariable, Pantograph
cannot catch the issue until the end of the proof. Anything that cannot be
expressed in Lean's type system (e.g., Homotopy Type Theory (HoTT)) also cannot
be expressed in Pantograph. Due to the tight coupling of Pantograph with Lean's
internals, non-trivial engineering effort is required to update Pantograph when
Lean undergoes a major version change.}

\Rebuttal{C3}{Moreover, due to the user-defined nature of tactics, distributing
computation via pickling of objects in Pantograph is not trivial. For example, if two branches
of a proof executing on two different machines are concluded, Pantograph does not
handle the algorithmically difficult problem of uniting the two branches.}




\section{Evaluation}
\label{sec:eval}
In this section, we demonstrate that Pantograph has the necessary power to implement a
Draft-Sketch-Prove (DSP) solver based on GPT-4o (which, notably, is not tuned for the task) and
Aesop \citep{dsp2022} (mentioned in Section~\ref{sec:related}). \Rebuttal{A2}{DSP
works as follows: A formal mathematical problem is fed into a language model
instance. The language model outputs a proof skeleton outlining the major steps
of the formal proof. Then a proof automation tool, which does not use machine
learning, fills in the holes.}

We used both the GPT-4o~\cite{gpt4o} and the GPT-o1-preview~\citep{o1-preview}
language models, with parameters given in Table~\ref{table:Parameters for DSP},
and ran on
the theorem proving evaluation benchmark MiniF2F \citep{minif2f}.
Each individual experiment works as follows. The language model is
given the formal theorem statement from MiniF2F and is asked to generate a natural
language proof. Next, the same model is provided with this natural language
proof and asked to generate one or three formal proof sketches in Lean~4
(GPT-o1-preview is only asked to generate a single proof sketch, as it does not
yet support multiple sketches). These sketches may contain the
\lstinline{sorry} keyword. The sketches are then fed into Pantograph's
sorry-extraction command and turned into goals, which we try to solve
one by one.  To attempt to solve the goals, we use the following Lean~4
tactics as hammers: \lstinline{aesop} ~\citep{aesop},
\lstinline{simp}, and \lstinline{linarith} (from Mathlib \citep{mathlib2019}).

\begin{table}
	\centering
	\begin{tabular}{|l|r|}
		\hline
		\textbf{Parameter} & \textbf{Value} \\
		\hline
		Max tokens & $2048$ \\
		Top P & $0.95$ \\
		Temperature & $0.8$ \\
		\hline
	\end{tabular}
        \vspace*{1em}
	\caption{LLM parameters for DSP Experiment}
	\label{table:Parameters for DSP}
\end{table}
Having built the DSP solver, the results are shown Table~\ref{table:Result of running Lean
DSP experiments}. They are not state of the art but are better than expected since we conducted no training. The drafting feature of Pantograph made this easy to implement. We show results on both the Validation and Test subsets of
the MiniF2F benchmark set, both with one and three proof sketch(es).  We report
the overall success rate, and the average number of hammer invocations and the average
runtime per benchmark.  Our best configuration uses GPT-4o and three requested
sketches.  For this configuration, the DSP system out of the box successfully proved $28\%$ of
the theorems from MiniF2F~\citep{minif2f}.  Since Lean and Isabelle have
different type systems, we do not do a comparison with the Isabelle
implementation of DSP.
\Comment{BM}{Why is this true or why does this comment matter? Should we write a minor footnote to explain? Or why does this matter? For TACAS perhaps this comment doesnt matter but the FM experts have to comment, but an ML reviewer would mainly care that the performance/accuracy is different btw the two system and then choose that system instead of the other.}

\begin{table}
	\centering
	\begin{tabular}{|l|r|r|r|r|r|r|}
		\hline
		Benchmark Set & \multicolumn{3}{l|}{Validation}
		& \multicolumn{3}{l|}{Test } \\
		\hline
		Requested Sketches (Model) & 1 (4o) & 1 (o1) & 3 (4o) & 1 (4o) & 1 (o1) & 3 (4o) \\
		\hline
		Success Rate ($\%$) & $12.7$ & $10.9$ & $23.6$ & $14.7$ & $16.0$ & \textbf{28.4} \\
		Hammer Invocations  & $4.17$ & $5.38$ & $7.93$ & $4.46$ & $5.72$ & $7.34$ \\
		Runtime (s)         & $17.25$ & $73.23$ & $23.98$ & $28.39$ & $88.38$ & $36.41$ \\
		\hline
	\end{tabular}
        \vspace*{1em}
	\caption{DSP's proof success rate (in \%) using the Pantograph interface on the MiniF2F formal theorem proving benchmark. 
	We used GPT-4o (labeled 4o) and o1-preview (labeled o1) for the DSP experiments.}
	\label{table:Result of running Lean DSP experiments}
\end{table}
\Comment{BM}{I dont think its clear what the table means. Without our zulip conversation I wouldnt have known that o1 wasnt able to getnerate multiple sketches}
\Comment{BM}{I dont think its clear which models generate what, perhaps we should include a model row (or an indication which accuracy corresponds to which model?)}
\Comment{BM}{Minor note, I personally prefer the \% to be present next to all number so I know immediately the units}
\Comment{BM}{I suggest we tell the reader explictly 100\% what the crips takeaway is from the figures. Right now I see counts, hammered goals etc and it seems a little bit like random statistics, we shoudl interpert the data ourselves and give the direct takeway to the user}
\Comment{BM}{It is not clear what valid1, valid3, test1, test3 means. I suggest figures to be self contained, given some readers might skip directly to the figures to see the evidence/takeways/main msg of our paper.}

\begin{figure*}[t!]
    \centering
    \begin{subfigure}[t]{0.5\textwidth}
        \centering
        \includegraphics[width=1\textwidth]{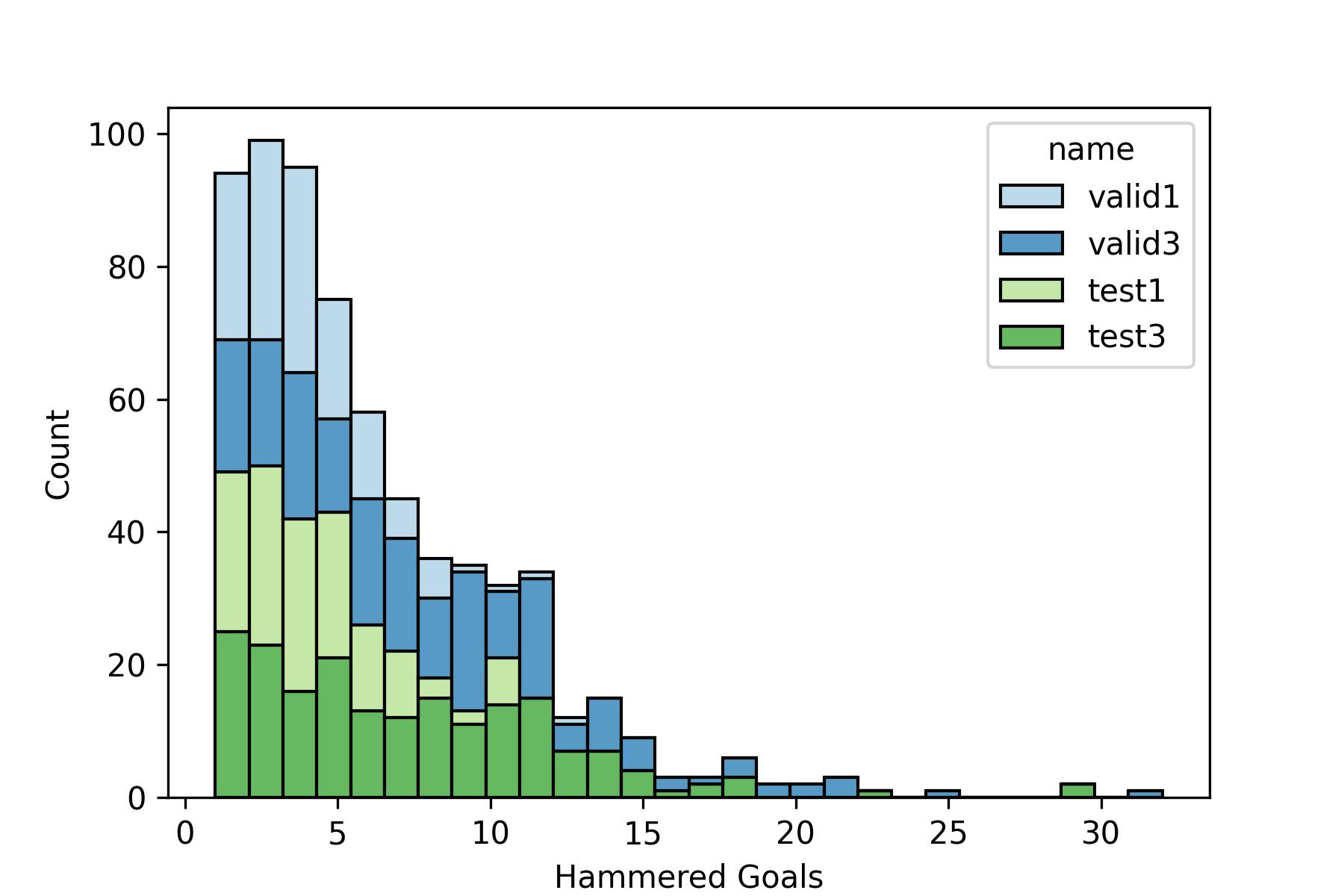}
        \caption{Number of goals generated for the hammer tactic}
    \end{subfigure}%
    \begin{subfigure}[t]{0.5\textwidth}
        \centering
        \includegraphics[width=1\textwidth]{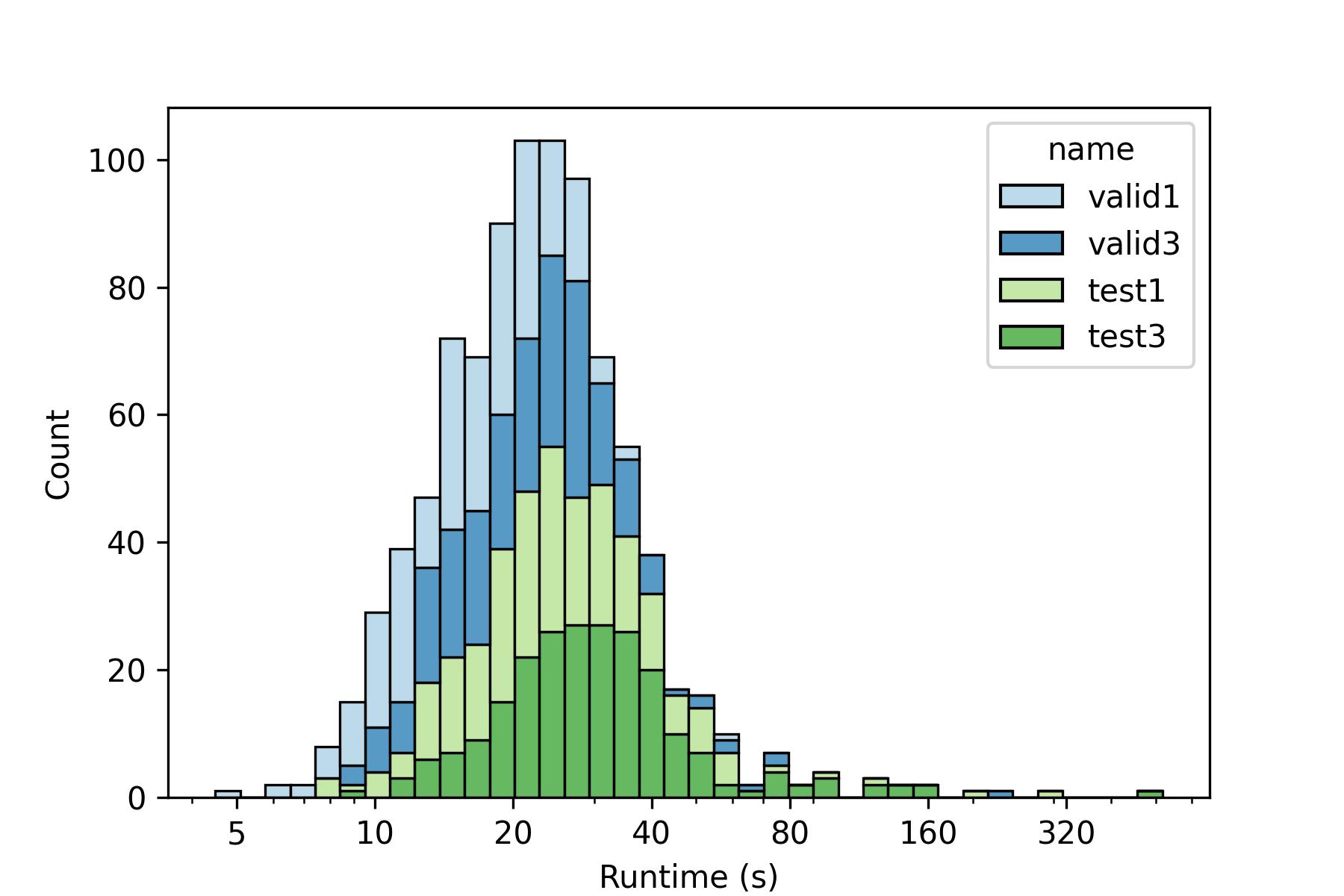}
        \caption{Total Runtime}
    \end{subfigure}
    \caption{Hammer invocations and runtimes of DSP on the validation and test
sets of MiniF2F using the GPT-4o model. The name of the legend refers to the
dataset split (validation or test) and the number of sketches used to solve
the dataset split.}
	\label{fig:Graph of hammer invocations and runtimes}
\end{figure*}

Figure~\ref{fig:Graph of hammer invocations and runtimes} shows the result of
the DSP experiment on the validation and test sets of the MiniF2F dataset. We
plot the distribution of the number of hammer tactic invocations and the
distribution of runtimes.  The LLM nearly always outputs fewer than 10 goals per
sketch.  We also observe that running 3 sketches rather than 1 does not
dramatically increase the runtime, indicating that the main performance
bottleneck is the inference of the GPT-4o model.

\begin{figure*}[t!]
    \centering
    \begin{subfigure}[t]{0.5\textwidth}
        \centering
        \includegraphics[width=1\textwidth]{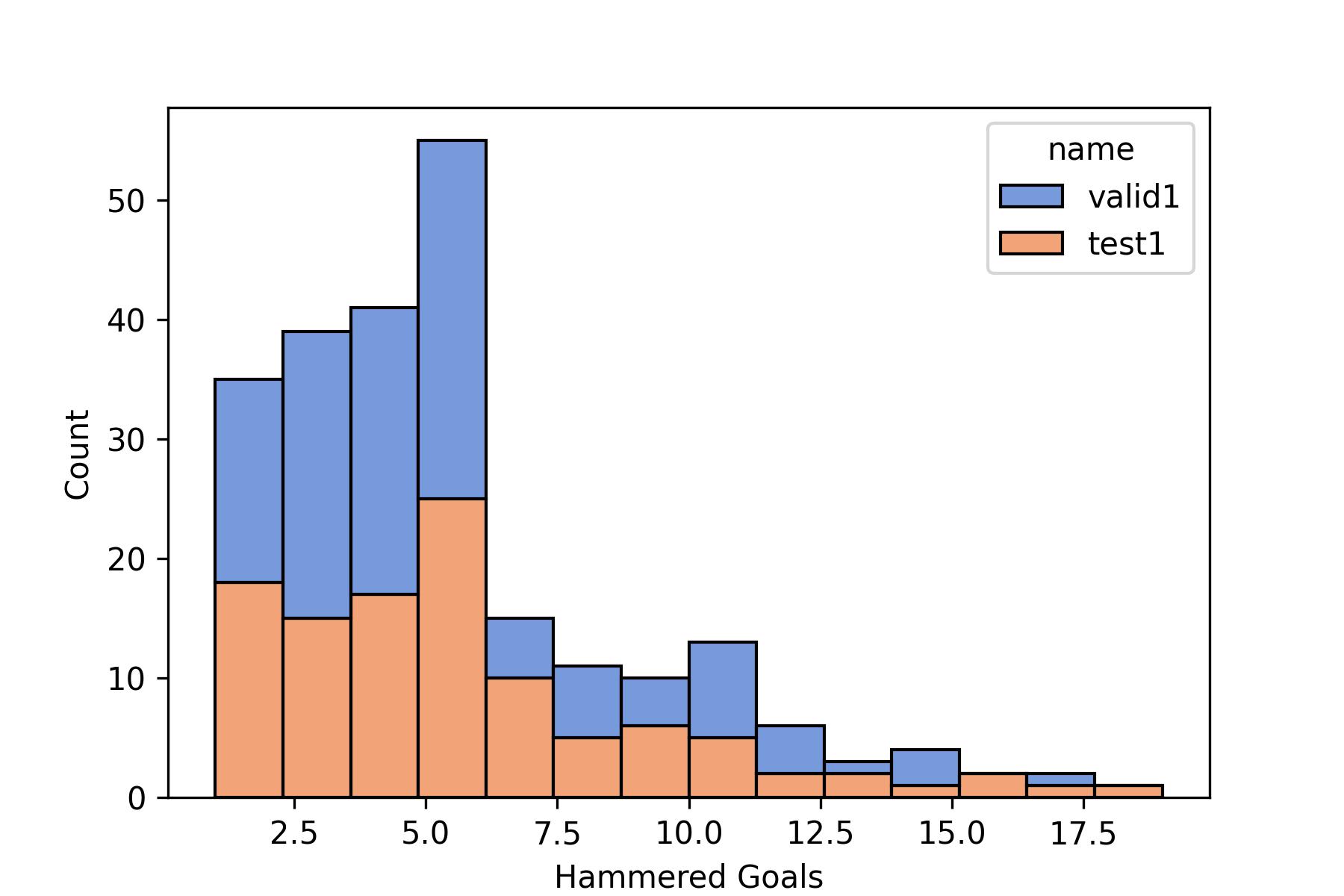}
        \caption{Number of goals generated for the hammer tactic}
    \end{subfigure}%
    \begin{subfigure}[t]{0.5\textwidth}
        \centering
        \includegraphics[width=1\textwidth]{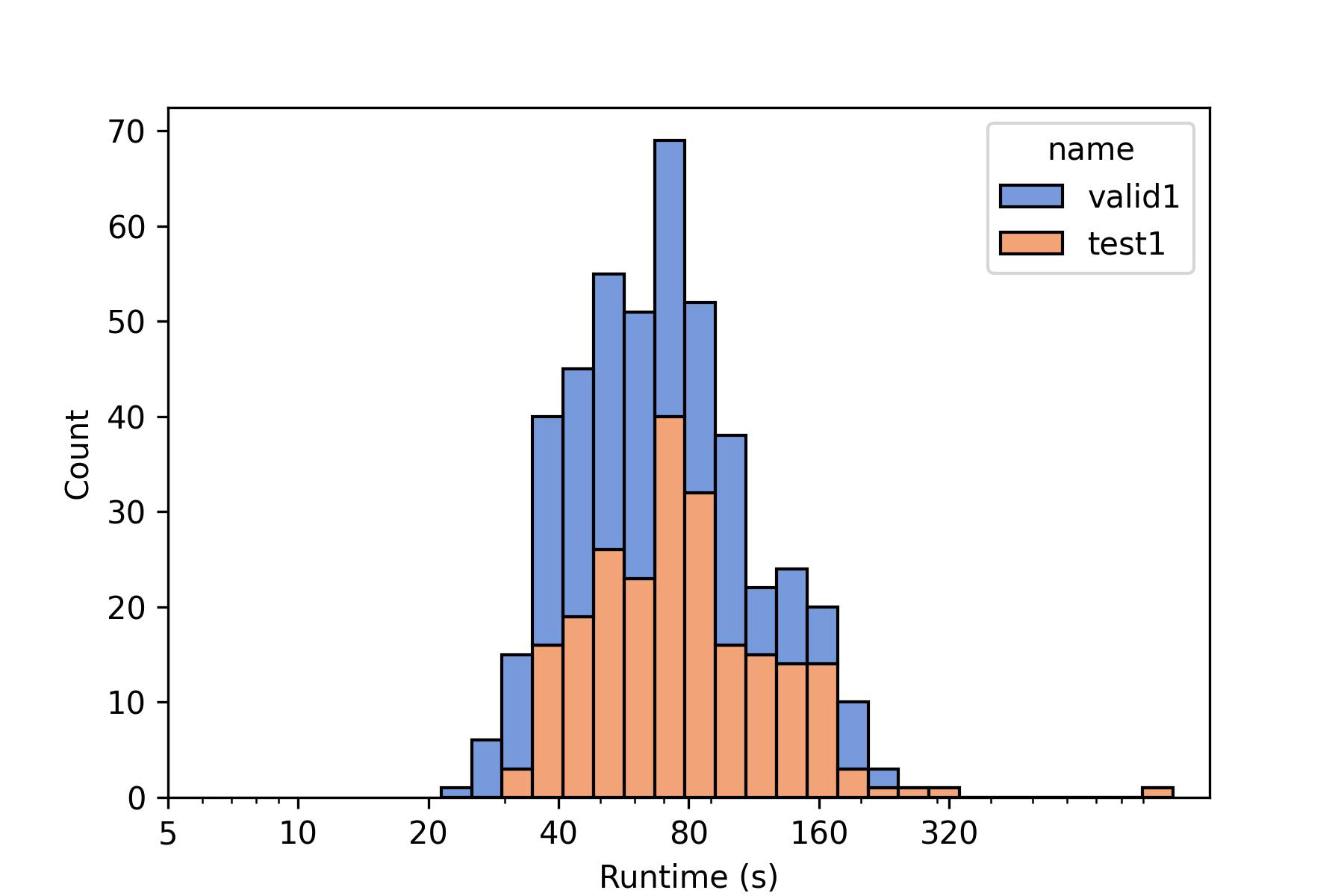}
        \caption{Total Runtime}
    \end{subfigure}
    \caption{Hammer invocations and runtimes of DSP on the validation and test
sets of MiniF2F using the o1-preview model. The name of the legend refers to the
dataset split (validation or test) and the number of sketches used to solve
the dataset split.}
	\label{fig:Result of o1-preview}
\end{figure*}

In Figure~\ref{fig:Result of o1-preview}, we show a similar plot for
the o1-preview model. This model cannot generate multiple
sketches. We observe that the runtime is much longer, likely due to the complex
inference mechanism of the o1-preview model. However the success rate is either
worse than or only marginally better than the GPT-4o model.

To our knowledge, these results represent the first successful implementation of
DSP in Lean~4.  We expect that the performance can be improved significantly by
tuning parameters or using more refined models, but this work provides a
baseline that can be built on and compared to in future work.


\section{Conclusion}
\label{sec:conclusion}
In this work, we introduce Pantograph, a Machine-to-Machine interaction library
for Lean~4.  We compare its features against existing tools used for training
machine learning models for theorem proving, and we provide a list of its novel
features. We also illustrate an application by implementing the first Lean~4
implementation of the Draft-Sketch-Prove approach.

In future work, we plan to use Pantograph to build and train various machine
learning approaches for theorem proving.  We also expect and hope that others will use
it in interesting and novel ways and that these use cases will provide feedback for
additional improvements and extensions of Pantograph.

\Rebuttal{A1}{Our evaluation also demonstrates one way that formal tools like
Lean can be used to address potential harm from 
Language Models such as the one used in the evaluation
section. Language models, though powerful, still face the problem of hallucination and generation
of illogical results. These can be mitigated by applying formal techniques
to the results produced of language models.  The draft-sketch-prove experiment is an
instance of this general idea, where proof automation formally checks the
potentially incorrect result generated by an LLM.  In the future, Pantograph
could be used for other hybrid reasoning approaches combining generative AI and
formal reasoning.}


\subsection*{Acknowledgements}

We thank Abdalrhman Mohamed for his input on Lean~4's expression system. We
also thank Lean~4 Zulip chat users who answered our development questions. We
thank Dr. Kim Morrison for the lean-training-data repository and examples of
interacting with Lean~4's front end. We thank Dr. David Dill for his feedback in
proofreading the paper.  This work was funded in part by a gift from Amazon Web
Services and by Centaur: the Stanford Center for Automated Reasoning.

\bibliographystyle{splncs04}
\bibliography{references}


\section*{Appendix}
\lstset{
	escapeinside={(**}{**)},
}

The prompt for the Draft part of the DSP experiment is
\begin{lstlisting}[language=none]
Draft an informal solution similar to the one below. The informal
solution will be used to sketch a formal proof in the Lean 4 Proof
Assistant.
Here are some examples:

Informal:
(*### Problem\n\n
[...nl/i problem text...]\n\n
### Solution\n\n
[...nl/i solution/draft text...]\n\n
*)\n\n

Informal:
(*### Problem\n\n
{nl_problem}
### Solution\n\n
[...Model Completion...]
\end{lstlisting}
The prompt for the Sketch part is
\begin{lstlisting}[language=none]
[... Translate informal draft to a formal sketch in Lean 4. Here are some examples: ...]
Informal:\n
(*### Problem\n\n
[...nl/i problem text...]\n\n
### Solution\n\n
[...nl/i solution/draft text...]\n\n
*)\n\n
Formal:\n
[...fl/i problem text...]
[...fl/i partial sketch text...]
\n\n

Informal:\n
(*### Problem\n\n
{nl_problem}
### Solution\n\n
{nl_solution}
*)\n\n
Formal:\n
{fl_problem}
[...Model Completion...]
\end{lstlisting}

\end{document}